\documentclass[final,5p,times,twocolumn]{elsarticle}

\usepackage{amssymb}
\usepackage{amsmath}
\usepackage{lineno}
\usepackage{amsfonts, amsthm, bm}
\usepackage{algorithm}
\usepackage{algpseudocode}
\usepackage{appendix}
\usepackage{url}
\usepackage{bm}

\journal{Elsevier Journal}

\begin{document}

\begin{frontmatter}



\title{Making the RANMAR pseudorandom number generator in LAMMPS up to four times faster, with an implementation of jump-ahead}


%% \author{} %% Author name

%% \affiliation{organization={},%Department and Organization

\author[label1]{Hiroshi Haramoto\corref{cor1}}
\affiliation[label1]{organization={Center for Data Science, Ehime University},
            addressline={3 Bunkyocho},
            city={Matsuyama},
            postcode={790-8577},
            state={Ehime},
            country={Japan}}
\ead{haramoto@ehime-u.ac.jp}
\cortext[cor1]{Corresponding Author}

\author[label2]{Kosuke Suzuki}
\affiliation[label2]{organization={Faculty of Science, Yamagata University},
            addressline={1-4-12 Kojirakawa-machi},
            city={Yamagata},
            postcode={990-8560},
            state={Yamagata},
            country={Japan}}
\ead{kosuke-suzuki@sci.kj.yamagata-u.ac.jp}

%% Abstract
\begin{abstract}
  Massively parallel molecular simulations require pseudorandom number streams
  that are provably non-overlapping and reproducible across
  thousands of compute units in parallel computing environments.
  In the widely used LAMMPS package, the standard RANMAR generator
  lacks a mathematically exact mechanism to jump ahead;
  distinct seeds are typically assigned instead, which does not ensure disjoint streams.
  We introduce a mathematically exact jump-ahead extension for RANMAR in LAMMPS.\@
  In practice, a single random sequence can be partitioned into consecutive,
  non-overlapping blocks of length $J$, with one block assigned
  to each compute unit under formal non-overlap guarantees.
  In our approach, we develop an algebraic reformulation
  that enables efficient jump-ahead even for very large $J$
  by casting state advancement into polynomial computations
  over finite residue rings while keeping memory small.
  We implement the extension in C++ using Number Theory Library (NTL)
  and integrate it into LAMMPS without altering user workflows.
  Beyond enabling exact partitioning, converting the 24-bit floating-point
  recurrence to an equivalent 24-bit integer recurrence accelerates generation itself:
  across diverse CPUs, generation is approximately two to four times faster
  than the floating-point baseline. Computing very large jumps (e.g., $J \approx 2^{120}$)
  remains practical.
\end{abstract}


% \begin{highlights}
% \item Mathematically exact jump-ahead for RANMAR integrated into LAMMPS
% \item Algebraic reformulation enables efficient jump-ahead for large sequences
% \item Non-overlapping, reproducible streams across many compute units
% \item 24-bit integer reformulation yields two to four times faster generation
% \end{highlights}

%% Keywords
\begin{keyword}
  random number generation \sep jump-ahead \sep RANMAR \sep LAMMPS \sep
  polynomial arithmetic \sep parallel computing


\end{keyword}

\end{frontmatter}


\section{Introduction}

The Large-scale Atomic/Molecular Massively Parallel Simulator (LAMMPS) is
a classical molecular dynamics code \cite{LAMMPS}. 
LAMMPS can be built and run on a single laptop or desktop machine, but it is also designed for parallel computers.
%% designed for parallel computing environments, though it can also be built and run on single laptops or decuyeretsktop machines. 
To support stochastic simulations, LAMMPS includes built-in facilities for pseudorandom number generation (PRNG),
and one of its standard generators is RANMAR proposed by Marsaglia, Zaman, and Tsang
\cite{MARSAGLIA199035}.
RANMAR is a combined generator consisting of a lagged Fibonacci generator and an arithmetic sequence generator.
The lagged Fibonacci generator ensures a sufficiently long period,
and the arithmetic sequence generator improves empirical statistical quality.
To ensure identical sequences across the heterogeneous CPU architectures prevalent
at the time of its proposal,
RANMAR was designed to generate 24-bit precision floating-point pseudorandom numbers.

With the development of hardware, machines with multiple compute units (e.g., GPUs) have become widespread,
leading to research on PRNGs suitable for parallel environments.
For example, Kokubo et al.\ parallelized the implementation of RANMAR in LAMMPS using SIMD and OpenMP
and reported approximately 1.5 times faster generation \cite{Tatsunobu-KOKUBO20192019-0008};
this approach illustrates the use of multiple compute units to produce a single pseudorandom sequence.
Regarding the parallelization of broader classes of PRNGs, L’Ecuyer et al.\ discuss requirements and methods for GPUs \cite{LECUYER20173},
and many implementations of commonly used generators on GPUs have been studied 
\cite{ALURU1992839, BRADLEY2011231, DEMCHIK2011692, PHILLIPS20117191}. 
%LECUYER20173,doi:10.1287/ijoc.1070.0251,10.1007/978-3-540-85912-3_26,JAMES1990329,PHILLIPS20117191,CHETRY201964,MAKINO19941357,}

Alongside such ``single-sequence, multi-unit'' strategies, an equally important technique
for parallel PRNGs is ``splitting a single RNG into equally spaced blocks (streams)'' via a jump-ahead.
In this approach, the output sequence, whose length equals the maximal period, is partitioned
into equal-length subsequences, and each compute unit is assigned a distinct initial state
corresponding to its designated subsequence. 
Specifically, let $J$ be a positive integer larger than the number of random variates consumed by a single unit.
If the first unit is initialized with state $s_{0}$, the second unit is initialized with state $s_{J}$, the state after $J$ steps from $s_0$,
the third unit is initialized with $s_{2J}$, and so on. This avoids overlap among streams.
The method is highly reproducible because one needs only the initial state $s_0$ and the jump size $J$
to reconstruct all stream starts.
Moreover, for certain classes of PRNGs, the structure of the underlying generator (e.g., lattice geometry)
allows theoretical assessment of inter-stream correlation \cite{10.1145/321371.321379},
enabling more stringent quality evaluation than statistical testing alone.

As argued above, non-overlap across compute units is essential to avoid spurious correlations and to ensure the reliability of large-scale simulations. By contrast, the current LAMMPS implementation of RANMAR lacks a jump-ahead capability and instead employs James’s initialization scheme, which refines the original Marsaglia--Zaman--Tsang procedure \cite{JAMES1990329} and assigns distinct initial states to each compute unit.
James claimed that, for each nonnegative integer seed up to \(9\times 10^{8}\), this method yields a distinct, non-overlapping sequence of substantial length (on the order of \(10^{30}\)).
However, to the best of our knowledge, no mathematical proof of this claim has been provided; in particular, Marsaglia et al.\ presented the initialization merely as an example and did not address the property asserted by James.
Consequently, there is no guarantee that independently seeded runs with the RANMAR implementation in LAMMPS are overlap-free.

To ensure overlap-free parallel streams in practice, it is therefore necessary to implement a jump-ahead for RANMAR.\@
Because RANMAR is a combined generator consisting of a lagged Fibonacci generator and an arithmetic sequence generator, we first consider jump-ahead methods for the lagged Fibonacci generator; the jump-ahead for the arithmetic-sequence generator can be reduced to modular arithmetic.
Chetry et al.\ introduced the notion of a \emph{jump index} for lagged Fibonacci generators \cite{CHETRY201964}. However, determining whether a jump index exists and, if so, whether its magnitude is large enough to avoid inter-stream overlap reduces to a kind of discrete logarithm problem. This task is generally intractable. In addition, the associated jump size is fixed, preventing arbitrary $J$-step jumps and making it difficult to shift stream offsets for comparative benchmarking under otherwise identical settings.

Makino proposed a matrix-based jump-ahead method for lagged Fibonacci generators \cite{MAKINO19941357}. If the recurrence has order $r$, the method computes an arbitrary $J$-step jump in time $O(r^{2}\log J)$ using an array storing $3r$ values. Although it applies only to the lagged Fibonacci part of the combined generator, it provides a useful benchmark for developing a jump-ahead for RANMAR.\@

In this article, we adapt a jump‑ahead technique originally developed for $\mathbb{F}_{2}$‑linear generators \cite{10.1007/978-3-540-85912-3_26, doi:10.1287/ijoc.1070.0251} to RANMAR, by formulating the next‑state computation as polynomial arithmetic over the residue ring $\mathbb{Z}/M\mathbb{Z}$, where $M=2^e$ is the modulus associated with the $e$‑bit precision. To implement a mathematically exact jump-ahead for RANMAR, we convert its 24‑bit floating‑point recursion to an equivalent 24‑bit integer recursion. This conversion guarantees that all coefficients in the ensuing algebra are 
integer residues, which is essential for an exact polynomial formulation of the transition function.

Using classical polynomial arithmetic, the resulting jump‑ahead runs in time $O(r^{2}\log J)$, comparable to Makino’s bound, while requiring only an array of length $r+1$ for working memory.
In practice, polynomial computations are carried
out using standard fast polynomial arithmetic, implemented in NTL \cite{NTL-SHOUP}. The formal 
setup appears in Section~3, and the timings are summarized in Table~2.
A further advantage, while not our primary objective, 
is that the integer representation incidentally speeds up random number generation by roughly 
a factor of two to four on modern CPUs. 

The remainder of the paper is organized as follows. Section~2 describes the RANMAR algorithm following Marsaglia et al.\ and formulates jump-ahead via a finite-state automaton. Section~3 explains the conversion from floating-point to integer arithmetic and the attendant polynomial computations. Section~4 details the implementation and reports generation-speed comparisons and jump-ahead timings. Section~5 concludes with perspectives and additional remarks. Appendix~A gives an integer-arithmetic re-implementation of the James-style initialization used in LAMMPS.\@

%LAMMPS is implemented in C++, thus we provide a C++ implementation of RANMAR
%that performs pseudorandom number generation and initialization
%using modular arithmetic. The code is available at \url{https://github.com/***}.
%We also provide a C++ implementation of the jump-ahead computation at the same site;
%as noted earlier, the jump-ahead relies on the NTL C++ number-theory library.

\section{Mathematical Structure of RANMAR}
\label{sec:intro_ranmar}
In this section, we summarize RANMAR using the notation of
the original paper \cite{MARSAGLIA199035}.
Furthermore, to make the mathematical structure of RANMAR explicit,
we formalize it as a finite-state automaton.

Recall that RANMAR is a combined generator consisting of two sequence generators.
One is a lagged Fibonacci generator, and the other is an arithmetic sequence generator.

For a lagged Fibonacci generator, we define the binary operation $x \bullet y$
for $x, y \in [0,1)$ by
$$
x \bullet y := x-y \pmod{1} = \begin{cases} x-y & (x \geq y) \\ x-y+1 & (x < y) \end{cases}.
$$
Let $r$ be a positive integer, and $s$ be a positive integer less than $r$.
For initial values $(x_1, x_2, \ldots, x_r) \in [0,1)^r$,
the lagged Fibonacci generator provides the sequence by the recurrence
$$
x_n : = x_{n-r} \bullet x_{n-s} \quad (n=r+1,r+2,\ldots).
$$

Next, we consider an arithmetic sequence generator.
Let $M$ be a positive integer, $m$ be a prime number less than $M$,
and $d$ be a constant less than $m$. 
We define the binary operation $c \circ d$ for $c, d \in [0,m/M)$ by
$$
  c \circ d := \begin{cases} c-d & (c \geq d) \\ c-d+m/M & (c < d) \end{cases}. 
$$
For an initial value $c_1 \in [0,m/M)$, the arithmetic sequence generator
provides the sequence by the following recursion:
$$
c_n :=  c_{n-1} \circ d \quad (n=2,3,\ldots). 
$$

Combining these values $x_1, x_2, \ldots$ and $c_1, c_2, \ldots$,
RANMAR generates the real numbers 
$$
x_1 \bullet c_1, \quad x_2 \bullet c_2, \quad \ldots, \quad x_n \bullet c_n, \quad \ldots
$$
as pseudorandom numbers from the initial values $x_1, x_2, \ldots, x_n \in [0,1)$
and $c_1 \in [0,m/M)$.

%Due to the architecture of CPUs at the time of \cite{MARSAGLIA199035},
Marsaglia et al.\ assumed the above numbers $x_i$, $c_i$,
and $x_i \bullet c_i$ with 24-bit precision, for $i=1,2,\ldots$.
Thus, they proposed the parameters $(r,s)=(97,33)$ for the lagged Fibonacci generator, 
$(M,m,d)=(2^{24}, 2^{24}-3, 7654321/2^{24})$ and the initial value $c_1=362436/2^{24}$
for the arithmetic sequence generator; the period of the resulting generator is
approximately $2^{144}$. 
LAMMPS adopts those parameters and initial values 
in the current version of the code \texttt{random\char`_mars.cpp}.

To prepare a jump-ahead of RANMAR, we represent it by a finite automaton.
A finite automaton of the lagged Fibonacci generator is as follows.
The state space is $[0,1)^r$, and the state transition function $f_1$ is 
\begin{align*}
\label{eq:state_transition}
&f_1: [0,1)^r \to [0,1)^r; \\
&     \qquad (x_1, x_2, \ldots, x_r) \mapsto
    (x_2, \ldots, x_r, x_1 \bullet x_{r-s+1}).
\end{align*}
The output space of the lagged Fibonacci generator is $[0,1)$,
and the output function $o_1$ is
$$
o_1 : [0,1)^r \to [0,1); \quad (x_1, \ldots, x_r) \mapsto x_1.
$$
    
We subsequently consider a finite automaton of the arithmetic sequence generator.
The state space is $[0,m/M)$, and the state transition function $f_2$ is
$$
f_2: [0,m/M) \to [0,m/M); \quad c \mapsto c \circ d. 
$$
The output space of the arithmetic sequence generator is $[0,1)$,
and the output function $o_2$ is the inclusion map,
i.e., $o_2(c) = c \in [0, 1)$ for any $c \in [0,m/M)$. 

We now describe a finite automaton representation of RANMAR and its jump-ahead.
The state space of RANMAR $S$ is 
$S:=[0,1)^r \times [0,m/M)$, and the state transition function is 
\begin{align*}
& f: S \to S; \\
  & \quad  (x_1, \ldots, x_r, c)  \\
  & \quad \mapsto (f_1(x_1, \ldots, x_r), f_2(c))
= (x_2, \ldots, x_{r}, x_1 \bullet x_{r-s+1}, c \circ d).
\end{align*}
The output space of RANMAR is $[0,1)$, and the output function $o$ is
$$
o: S \to [0,1); \quad (x_1, \ldots, x_r, c) \mapsto x_1 \bullet c
 = o_1(x_1, \ldots, x_r) \bullet o_2 (c). 
$$  

For a positive integer $J$, let $f^J$, $f_1^J$ and $f_2^J$ denote
the $J$-th iteration of the function $f$, $f_1$ and $f_2$.
The $J$-th state from the initial state $(x_1, x_2, \ldots, x_r, c_1) \in S$ 
is 
$$
f^J (x_1, \ldots, x_r, c) = (f_1^J(x_1, \ldots, x_r), f_2^J(c)), 
$$
and hence we have
\begin{align*}
o(f^J(x_1, \ldots, x_r, c)) = o_1(f_1^J(x_1, \ldots, x_r)) \bullet o_2(f_2^J(c)). 
\end{align*}
Thus the $J$-th output of RANMAR depends only on
the $J$-th state of the lagged Fibonacci generator and
the $J$-th state of the arithmetic sequence generator. 
This shows that a jump-ahead for RANMAR can be implemented by performing a jump-ahead independently for the lagged Fibonacci generator and the arithmetic-sequence generator.

\section{Modular Arithmetic Representation}
RANMAR in LAMMPS is implemented using floating-point operations as described in the previous section.
%The reason why Marsaglia et al.\ adopted the description was to generate the same pseudorandom numbers
%on various computers at the time, and they also assumed the use of Fortran.
On the other hand, it is typical to describe lagged Fibonacci generators using modular arithmetic
\cite{Knuth:1997:ACP:270146}. Marsaglia et al.\ also provided an equivalent definition
using modular arithmetic briefly, in addition to the floating-point operation-based definition
of RANMAR mentioned above.
Accordingly, we will provide a detailed explanation of RANMAR using modular arithmetic.

Hereafter, let $e$ denote the precision of real numbers and let $M$, 
one of the parameters of the arithmetic sequence generator, be $2^e$;
as we mentioned before, RANMAR adopts $e=24$.

\subsection{Jump-ahead for the lagged Fibonacci generator}\label{sec:jump-fib}

In the algebra system $(\mathbb{R}/\mathbb{Z}, +)$, where the operation $+$ is the canonical 
addition over $\mathbb{R}/\mathbb{Z}$, the binary operation $\bullet$ coincides
with the subtraction of this group. 

Since $e$ is the precision of real numbers, the set of all real numbers considered hereafter is
$$
G_e:=\left\{\left. \sum_{i=1}^e \frac{a_{-i}}{2^i} \,\,  \right|
  \,\, a_{-1}, a_{-2} \ldots, a_{-e} \in \{0, 1\} \right\} \subset [0,1).
$$
Hence, the bijection
$$
\sigma : G_e \to \mathbb{Z}/2^e\mathbb{Z}; \quad
\sum_{i=1}^e \frac{a_{-i}}{2^i} \mapsto 2^e\sum_{i=1}^e \frac{a_{-i}}{2^i} = 
\sum_{i=1}^{e} a_{-i} 2^{e-i}
$$
gives a group isomorphism between the additive groups 
$\left(G_e, +\right)$ and $(\mathbb{Z}/2^e\mathbb{Z}, +)$,
where $\mathbb{Z}/2^e\mathbb{Z}$ is the residue group modulo $2^e$.  

Let $F_1$ be the map 
\begin{align*}
  & F_1: (\mathbb{Z}/2^e\mathbb{Z})^r \to (\mathbb{Z}/2^e\mathbb{Z})^r; \\
  & \qquad (u_1, u_2, \ldots, u_r) \mapsto (u_2, \ldots, u_r, u_{1}-u_{r-s+1}),  
\end{align*}
which corresponds to the state transition function 
of the lagged Fibonacci generator $f_1$. 
That is, for any positive integer $J$, 
we have the equality $f_1 = \sigma^{-1} \circ F_1 \circ \sigma$, 
and thus $f_1^J = \sigma^{-1} \circ F_1^J \circ \sigma$.
This indicates that jumping ahead over $\mathbb{Z}/2^e\mathbb{Z}$ enables
jumping ahead over $G_e$. 

Let $A$ be the $r \times r$ matrix
$$
A=\begin{pmatrix} 0 & \cdots & 0 & a_0
\\ & I_{r-1} & & \begin{matrix} a_1 \\ \vdots \\ a_{r-1} \end{matrix}\end{pmatrix},
$$
where $I_{r-1}$ is the identity $(r-1)\times(r-1)$ matrix, and 
$a_0=1$, $a_{r-s}=-1$ and $a_i = 0$ for $i \neq 0, r-s$.
The state transition function $F_1$ can be written by
$$
F_1(u_1, u_2, \ldots, u_r) = (u_1, u_2, \ldots, u_r)A
$$
for $(u_1, u_2, \ldots, u_r) \in\left(\mathbb{Z}/2^e\mathbb{Z}\right)^r$.

The characteristic polynomial of $A$ is  
$$
\varphi(t) := \det(tI_r - A) = t^{r} + t^{r-s} - 1. 
$$
By polynomial division, there exists a polynomial
$P_J(t) \in (\mathbb{Z}/2^e\mathbb{Z})[t]$ 
with $\deg (P_J(t))< \deg(\varphi(t)) = r$ satisfying
$$
t^J \equiv P_J(t) \pmod{\varphi(t)}.
$$ 
According to the Cayley--Hamilton theorem \cite{lang02},
we have $A^J = P_J(A)$ and hence
\begin{align*}
  & (u_{J+1}, u_{J+2}, \ldots, u_{J+r}) \\
  & \quad = F_1^J(u_{1}, u_{2}, \ldots, u_{r}) \\
  & \quad = (u_{1}, u_{2}, \ldots, u_{r})A^J 
  = (u_{1}, u_{2}, \ldots, u_{r})P_J(A).
\end{align*}

Since $P_J(t) \in (\mathbb{Z}/2^e\mathbb{Z})[t]$, all coefficients lie in $\{0, 1, \ldots, 2^e-1\}$, enabling efficient computation of  $t^J \pmod{\varphi(t)}$.
Assume $P_J(t) = \sum_{i=0}^{r-1} b_i t^i$. For a state 
$\bm{u} = (u_1, u_2, \ldots, u_r) \in (\mathbb{Z}/2^e\mathbb{Z})^r$, 
the state $\bm{u}P_J(A)$ can be effectively computed by the Horner method:
$$
\bm{u}P_J(A) = ( \cdots ((b_{r-1}\bm{u}A+b_{r-2}\bm{u})A+b_{r-3}\bm{u})A + \cdots +
b_1\bm{u})A + b_0\bm{u}. 
$$

\subsection{Jump-ahead for the arithmetic sequence generator}
We next consider the jump-ahead for the arithmetic sequence generator.
Assume that the parameter $d$ and the initial value $c_0$ in the arithmetic sequence generator are rational numbers with a denominator of $2^e$.
That is, there exist integers $d'$ and $c'_0$ satisfying $d=d'/2^e$ and $c_0 = c'_0/2^e$.
By the definition of the recurrence, all terms in the sequence $c_0, c_1, \ldots$
are also rational numbers with a denominator of $2^e$.
Let $c'_i = c_i 2^e$ (i.e., the integer numerator of $c_i$).
Then, similar to Section~\ref{sec:jump-fib}, the recurrence relation for the numerators is:
\[
c'_{i+1} = (c'_i - d') \bmod m,
\]
where $A \bmod B$ denotes the remainder of $A$ divided by $B$.
Thus, jumping ahead $J$ steps to find the term $c_J$ can be computed as:
\[
c_J = \dfrac{(c'_0 - J \cdot d') \bmod m}{2^e}
\]
In computing this value, we only need $J \bmod m$, not $J$ itself.
Thus we can compute $c_J$ efficiently even if $J$ is extremely large (e.g., $J \gg 2^{60}$). 

\section{Implementations and time comparisons}
In this section, we show the implementation of RANMAR using modular arithmetic, 
compare the generating speed and the computation time of the jump-ahead. 

In order to clarify the improvement, we show a pseudocode of RANMAR in LAMMPS
in Algorithm \ref{alg:ranmars_original}.
The current implementation of RANMAR requires five if-statements.

\begin{algorithm}[ht]
\caption{the current implementation of RANMAR in LAMMPS}
\label{alg:ranmars_original}  
\begin{algorithmic}
\Require real numbers $x_1, \ldots, x_r, c$
\Ensure a real number $y$
\State $\triangleright$ The state transition of the lagged Fibonacci generator.
\State $y \gets x_{i}-x_{j}$
\If{$y < 0$}
\State $y \gets y+1$
\EndIf
\State $x_i \gets y$
\State $i \gets i-1$
\If{$i = 0$}
\State $i \gets 97$
\EndIf
\State $j \gets j-1$
\If{$j = 0$}
\State $j \gets 97$
\EndIf \\
\State $\triangleright$ The state transition of the arithmetic sequence generator.
\State $c \gets c-d$  
\If{$c < 0$}
\State $c \gets c + m/M$
\EndIf \\
\State $\triangleright$ Combine two real numbers into a single value.
\State $y \gets y-c$ 
\If{$y < 0$} 
\State $y \gets y + 1$
\EndIf
\end{algorithmic}
\end{algorithm}

Algorithm \ref{alg:ranmars_modular_arithmetic} shows an equivalent pseudocode
of RANMAR by using modular arithmetic.
Compared with Algorithm \ref{alg:ranmars_original}, modular arithmetic allows us
to reduce the number of if-statements from five to three.
If-statements are generally more costly than logical instructions,
so Algorithm \ref{alg:ranmars_modular_arithmetic} is expected to be faster.

LAMMPS is implemented in C++, and its handling of overflow in modular arithmetic
is well-suited for optimizing the performance and implementing a jump-ahead of RANMAR.\@
Throughout this section we assume that the machine word size $w$ satisfies $w \geq e$.
Contemporary CPUs typically adopt $w \in \{32,64\}$, so this is a natural assumption
for our setting with $e=24$.
In C++, arithmetic on $w$-bit unsigned types is defined modulo $2^{w}$;
consequently, for $u,v \in \{0,\ldots,2^{w}-1\}$ with $u<v$,
the subtraction returns the residue $u - v + 2^{w}$.
In particular, $u-v$ is always interpreted as a non-negative residue, which allows us to remove the first if-statement in Algorithm~1.

Moreover, since
$$
  (u - v) \bmod 2^{e}
  \;=\;
  \left( (u \bmod 2^{e}) - (v \bmod 2^{e}) \right) \bmod 2^{e},  
$$
the last if statement in Algorithm~1 can be replaced by masking the lower $e$ bits
with $2^{e}-1$ (e.g., \texttt{0x00FFFFFF} when $e=24$) and
then scaling by $2^{-e}$ (e.g., $2^{-24}$) to map the result to $[0,1)$.

\begin{algorithm}[ht]
\caption{A modular arithmetic implementation of RANMAR for LAMMPS}
\label{alg:ranmars_modular_arithmetic}  
\begin{algorithmic}
\Require unsigned integers of at least 24 bits $u_1, \ldots, u_r, v$
\Ensure a real number $y$
\State $\triangleright$ The state transition of the lagged Fibonacci generator.
\State $u \gets u_{i}-u_{j}$
\State $u_i \gets u$
\State $i \gets i-1$
\If{$i = 0$}
\State $i \gets 97$
\EndIf
\State $j \gets j-1$
\If{$j = 0$}
\State $j \gets 97$
\EndIf \\
\State $\triangleright$ The state transition of the arithmetic sequence generator.
\State $v \gets v+(m-d_0)$
\If{$v \geq m$}
\State $v \gets v - m$
\EndIf \\
\State $\triangleright$ Combine two integers into a single value, and convert it into
the floating-point number.
\State $y \gets ((u-v) \pmod{2^{24}}) / 2^{24}$
\end{algorithmic}
\end{algorithm}

We compare the generation speed of RANMAR on the following four environments: 
(1) Apple (MacOS 14.5, Apple Silicon M2 Pro, 32GB memory), 
(2) Raspberry Pi 4B (Linux Kernel 6.6.31+rpt-rpi-v8, Broadcom BCM2711 Cortex-A72
(1.5GHz), 4GB memory), (3) Ryzen (Linux Kernel 5.15.0-113-generic,
Ryzen Threadripper 3970X (3.7GHz), 128GB memory),
(4) Intel (Linux Kernel 5.15.0-91-generic, Intel i7-14700F (5.3GHz), 32GB memory).

Table~\ref{table:gen_speed} shows computation time of $10^9$ pseudorandom numbers
by the current RANMAR in LAMMPS (by real number computation)
and our proposal (by 24-bit modular arithmetic).
Modular arithmetic is approximately two to four times faster than the current RANMAR.\@

\begin{table}[ht]
  \centering
  \caption{Generating time of RANMAR (in seconds)}
  \label{table:gen_speed}
  \begin{tabular}{ccccc}
    \hline
     & Apple & ARM & Ryzen & Intel \\ \hline
    floating-point number & 3.30 & 16.60 & 8.45 & 3.31 \\
    modular arithmetic & 1.27 & 6.14 & 2.07 & 1.01 \\ \hline
  \end{tabular}
\end{table}

We then implement a jump-ahead by using the Number Theory Library (NTL) \cite{NTL-SHOUP},
which is a high-performance, portable C++ library providing
data structures and algorithms for manipulating polynomials over
$\mathbb{Z}/2^e\mathbb{Z}$.
Table~\ref{table:jump_speed} summarizes the speed of computing
$$
t^J \pmod{\varphi(t)}
$$ 
in the case $J=2^{64}-1$ and $J=2^{120}-1$; NTL can handle multi-precision integers. We choose exponents of the form $2^n-1$ because their binary representation consists entirely of ones, making the number of multiplications per bit identical for the binary-exponentiation routine.

\begin{table}[ht]
  \centering
  \caption{Time to compute $t^J \pmod{\varphi(t)}$ (in microseconds)}
  \label{table:jump_speed}
  \begin{tabular}{ccccc}
    \hline
    & Apple & ARM & Ryzen & Intel \\ \hline
    $J=2^{64}-1$ & 1347 & 10395 & 946 & 2702 \\
    $J=2^{120}-1$ & 2556 & 21411 & 1946 & 4697 \\ \hline
  \end{tabular}
\end{table}

The direct computation of the $J$-th iteration is impractical, 
but the jump-ahead described in Section 3 derives the $J$-th state
in a relatively short time. 
Because the degree of the characteristic polynomial $\varphi(t)$ 
of the lagged Fibonacci generator is $r=97$, 
the benefit of asymptotically faster multiplication algorithms (e.g., Karatsuba) over the classical quadratic algorithm is modest \cite{von_zur_Gathen_Gerhard_2013}. 
In practice, NTL selects appropriate algorithms for polynomial multiplication 
and division, enabling fast computation.

\section{Concluding Remarks}
In this paper, we present a jump-ahead method for RANMAR
pseudorandom number generator. Our method supports an arbitrary jump length
and thus guarantees non-overlapping subsequences.

If future implementations adopt a larger recurrence order $r$ 
for the lagged Fibonacci recurrence, the computational cost of evaluating 
the $J$-step jump-ahead scales with the complexity of the underlying polynomial arithmetic. 
With classical algorithms the cost is $O(r^{2}\log J)$; with faster polynomial
routines it becomes $O(r^{\log_{2}3}\log J)$ or better.
Since NTL automatically selects appropriate routines, our jump-ahead remains
computationally advantageous even for large $r$.

Note that LAMMPS has another standard PRNG called Park--Miller generator
\cite{10.1145/63039.63042} in \texttt{random\char`_park.cpp}; 
this generator is a variant of Lehmer generator \cite{MR0044899}.
The recurrence of Park--Miller generator is 
$$
X_{n+1} = 16807 X_{n} \pmod{2^{31}-1}. 
$$
Since the maximum period of this generator is $2^{31}-2$,
it is too short to consider a practical jump-ahead.
We do not consider the Park--Miller generator in this study. 
 
\appendix
\section*{Appendix A. Initialization of RANMAR in LAMMPS using modular arithmetic}

LAMMPS initializes the state space of RANMAR using the method described in \cite{MARSAGLIA199035}.
Since the initialization procedure in LAMMPS involves floating-point calculations,
we implement the equivalent initialization by modular arithmetic.
Algorithm \ref{alg:init_ranmars_modular_arithmetic} provides a pseudocode for the parameters currently implemented in LAMMPS and the specific feature of the C++ language. 

In order to distinguish between division in integer arithmetic and
division in floating-point computation, we write $\mathrm{Remainder}(a, b)$
(resp.\ $\mathrm{Quotient}(a,b)$) to denote the integer remainder
(resp.\ the integer quotient) when $a$ is divided by $b$ for $a, b \in \mathbb{Z}$.

We have numerically verified that the state produced by Algorithm~\ref{alg:init_ranmars_modular_arithmetic}, when converted back to floating‑point format, matches exactly the initialization sequence produced by the original floating‑point implementation in LAMMPS.\@  This ensures that our modular‑arithmetic 
reimplementation preserves the behavior of the existing code.

\begin{algorithm}[t]
\caption{Initializing Procedure of RANMAR generator}
\label{alg:init_ranmars_modular_arithmetic}  
\begin{algorithmic}
\Require an integer $s$ 
\State\Comment{$s$ is a seed satisfying $0 \leq s \leq 9 \times 10^8$.}
\Ensure $w$-bit unsigned integers $i, j, c_0, d_0, m, u_1, u_2, \ldots, u_{97}$ 
\State\Comment{The word size $w$ must be $w \geq 24$.}
\State $i_j \gets \mathrm{Quotient}(s-1, 30082)$
\State $k_l \gets s-1-30082 \times i_j$
\State $i \gets \mathrm{Remainder}(\mathrm{Quotient}(i_j, 177), 177) +2$
\State $j \gets \mathrm{Remainder}(i_j, 177)+2$
\State $k \gets \mathrm{Remainder}(\mathrm{Quotient}(k_l, 169), 178) + 1$
\State $l \gets \mathrm{Remainder}(k_l, 169)$
\For{$i_i \gets 1$ \textbf{to} $97$}
    \State $s \gets 0$
    \State $t \gets 2^{w-1}$
    \Comment{$2^{w-1}$ corresponds to $0.5$ in floating-point computation.}
    \For{$j_j \gets 1$ \textbf{to} 24}
    \State $m \gets \mathrm{Remainder}(\mathrm{Quotient}(i \times j, 179) \times k, 179)$
    \State $i \gets j$
    \State $j \gets k$
    \State $k \gets m$
    \State $l \gets \mathrm{Remainder}(53 \times l + 1, 169)$
    \If{$\mathrm{Remainder}(l\times m, 64) \geq 32$}
        \State $s \gets s+t$
    \EndIf
    \State $t \gets \mathrm{Quotient}(t, 2)$
    \State
    \Comment{It is equivalent to multiplying a floating point number by 0.5.}
    \EndFor
    \State $u_{i_i} \gets \mathrm{Quotient}(s, 2^{w-24})$
    \State
    \Comment{To convert $w$-bit integer to 24-bit integer, shift right by $w-24$ bits.}
\EndFor
\State $c_0 \gets 362436$
\State $d_0 \gets 7654321$
\State $m \gets 2^{24}-3$
\State $i \gets 97$
\State $j \gets 33$
\end{algorithmic}
\end{algorithm}

\section*{Code and data availability}
All source code are available at \url{https://github.com/HiroshiHaramoto/ranmar24-jumpahead}.

\section*{CRediT authorship contribution statement}
H. Haramoto: Conceptualization, Methodology, Formal analysis, Software,
Validation, Writing -- Original Draft. 
K. Suzuki: Software, Investigation, Writing -- Review  \& Editing.

\section*{Declaration of generative AI and AI-assisted technologies in the manuscript preparation process}
During the preparation of this work the authors used ChatGPT and Gemini in order to improving language and readability and refining the structure and organization of the paper. After using this tool/service, the authors reviewed and edited the content as needed and take full responsibility for the content of the published article.

\section*{Declaration of competing interest}
The authors declare that they have no known competing financial interests or personal relationships that could have appeared to influence the work reported in this paper.

\section*{Acknowledgments}
This work was supported by JSPS KAKENHI Grant Numbers 19K03450, 22K03415, 24K06857 and 25K07132.

%% \section{Example Appendix Section}
%% \label{app1}

%% Appendix text.

%%       \cite{<label>} ==> [1]

%%
%% Example citation, See \cite{lamport94}.

\bibliographystyle{elsarticle-num} 
\bibliography{haramoto01}

\begin{thebibliography}{10}
\expandafter\ifx\csname url\endcsname\relax
  \def\url#1{\texttt{#1}}\fi
\expandafter\ifx\csname urlprefix\endcsname\relax\def\urlprefix{URL }\fi
\expandafter\ifx\csname href\endcsname\relax
  \def\href#1#2{#2} \def\path#1{#1}\fi

\bibitem{LAMMPS}
A.~P. Thompson, H.~M. Aktulga, R.~Berger, D.~S. Bolintineanu, W.~M. Brown,
  P.~S. Crozier, P.~J. in't Veld, A.~Kohlmeyer, S.~G. Moore, T.~D. Nguyen,
  R.~Shan, M.~J. Stevens, J.~Tranchida, C.~Trott, S.~J. Plimpton, {LAMMPS} - a
  flexible simulation tool for particle-based materials modeling at the atomic,
  meso, and continuum scales, Comp. Phys. Comm. 271 (2022) 108171.
\newblock \href {https://doi.org/10.1016/j.cpc.2021.108171}
  {\path{doi:10.1016/j.cpc.2021.108171}}.

\bibitem{MARSAGLIA199035}
G.~Marsaglia, A.~Zaman, W.~{Wan Tsang},
  \href{https://www.sciencedirect.com/science/article/pii/016771529090092L}{Toward
  a universal random number generator}, Statistics \& Probability Letters 9~(1)
  (1990) 35--39.
\newblock \href {https://doi.org/https://doi.org/10.1016/0167-7152(90)90092-L}
  {\path{doi:https://doi.org/10.1016/0167-7152(90)90092-L}}.
\newline\urlprefix\url{https://www.sciencedirect.com/science/article/pii/016771529090092L}

\bibitem{Tatsunobu-KOKUBO20192019-0008}
T.~Kokubo, S.~Nagaoka, H.~Teramae, U.~Nagashima, {For a Super-High-Speed
  Cluster Type Parallel Computer (K Computer), Acceleration of General-Purpose
  Molecular Dynamics Program LAMMPS.}, Journal of Computer Chemistry, Japan
  18~(4) (2019) 169--175.
\newblock \href {https://doi.org/10.2477/jccj.2019-0008}
  {\path{doi:10.2477/jccj.2019-0008}}.

\bibitem{LECUYER20173}
P.~L’Ecuyer, D.~Munger, B.~Oreshkin, R.~Simard,
  \href{https://www.sciencedirect.com/science/article/pii/S0378475416300829}{Random
  numbers for parallel computers: Requirements and methods, with emphasis on
  {GPU}s}, Mathematics and Computers in Simulation 135 (2017) 3--17, special
  Issue: 9th IMACS Seminar on Monte Carlo Methods.
\newblock \href {https://doi.org/https://doi.org/10.1016/j.matcom.2016.05.005}
  {\path{doi:https://doi.org/10.1016/j.matcom.2016.05.005}}.
\newline\urlprefix\url{https://www.sciencedirect.com/science/article/pii/S0378475416300829}

\bibitem{ALURU1992839}
S.~Aluru, G.~Prabhu, J.~Gustafson,
  \href{https://www.sciencedirect.com/science/article/pii/016781919290030B}{A
  random number generator for parallel computers}, Parallel Computing 18~(8)
  (1992) 839--847.
\newblock \href {https://doi.org/https://doi.org/10.1016/0167-8191(92)90030-B}
  {\path{doi:https://doi.org/10.1016/0167-8191(92)90030-B}}.
\newline\urlprefix\url{https://www.sciencedirect.com/science/article/pii/016781919290030B}

\bibitem{BRADLEY2011231}
T.~Bradley, J.~{du Toit}, R.~Tong, M.~Giles, P.~Woodhams,
  \href{https://www.sciencedirect.com/science/article/pii/B9780123849885000164}{{Chapter
  16 - Parallelization Techniques for Random Number Generators}}, in: W.~mei
  W.~Hwu (Ed.), GPU Computing Gems Emerald Edition, Applications of GPU
  Computing Series, Morgan Kaufmann, Boston, 2011, pp. 231--246.
\newblock \href
  {https://doi.org/https://doi.org/10.1016/B978-0-12-384988-5.00016-4}
  {\path{doi:https://doi.org/10.1016/B978-0-12-384988-5.00016-4}}.
\newline\urlprefix\url{https://www.sciencedirect.com/science/article/pii/B9780123849885000164}

\bibitem{DEMCHIK2011692}
V.~Demchik,
  \href{https://www.sciencedirect.com/science/article/pii/S0010465510004868}{{Pseudo-random
  number generators for Monte Carlo simulations on ATI Graphics Processing
  Units}}, {Computer Physics Communications} 182~(3) (2011) 692--705.
\newblock \href {https://doi.org/https://doi.org/10.1016/j.cpc.2010.12.008}
  {\path{doi:https://doi.org/10.1016/j.cpc.2010.12.008}}.
\newline\urlprefix\url{https://www.sciencedirect.com/science/article/pii/S0010465510004868}

\bibitem{PHILLIPS20117191}
C.~L. Phillips, J.~A. Anderson, S.~C. Glotzer,
  \href{https://www.sciencedirect.com/science/article/pii/S0021999111003329}{{Pseudo-random
  number generation for Brownian Dynamics and Dissipative Particle Dynamics
  simulations on GPU devices}}, Journal of Computational Physics 230~(19)
  (2011) 7191--7201.
\newblock \href {https://doi.org/https://doi.org/10.1016/j.jcp.2011.05.021}
  {\path{doi:https://doi.org/10.1016/j.jcp.2011.05.021}}.
\newline\urlprefix\url{https://www.sciencedirect.com/science/article/pii/S0021999111003329}

\bibitem{10.1145/321371.321379}
R.~R. Coveyou, R.~D. Macpherson,
  \href{https://doi.org/10.1145/321371.321379}{{Fourier Analysis of Uniform
  Random Number Generators}}, J. ACM 14~(1) (1967) 100–119.
\newblock \href {https://doi.org/10.1145/321371.321379}
  {\path{doi:10.1145/321371.321379}}.
\newline\urlprefix\url{https://doi.org/10.1145/321371.321379}

\bibitem{JAMES1990329}
F.~James,
  \href{https://www.sciencedirect.com/science/article/pii/001046559090032V}{A
  review of pseudorandom number generators}, Computer Physics Communications
  60~(3) (1990) 329--344.
\newblock \href {https://doi.org/https://doi.org/10.1016/0010-4655(90)90032-V}
  {\path{doi:https://doi.org/10.1016/0010-4655(90)90032-V}}.
\newline\urlprefix\url{https://www.sciencedirect.com/science/article/pii/001046559090032V}

\bibitem{CHETRY201964}
M.~K. Chetry, S.~K. Bishoi, V.~Matyas,
  \href{https://www.sciencedirect.com/science/article/pii/S0166218X19303038}{{When
  Lagged Fibonacci Generators jump}}, Discrete Applied Mathematics 267 (2019)
  64--72.
\newblock \href {https://doi.org/https://doi.org/10.1016/j.dam.2019.06.022}
  {\path{doi:https://doi.org/10.1016/j.dam.2019.06.022}}.
\newline\urlprefix\url{https://www.sciencedirect.com/science/article/pii/S0166218X19303038}

\bibitem{MAKINO19941357}
J.~Makino,
  \href{https://www.sciencedirect.com/science/article/pii/0167819194900426}{{Lagged-Fibonacci
  random number generators on parallel computers}}, Parallel Computing 20~(9)
  (1994) 1357--1367.
\newblock \href {https://doi.org/https://doi.org/10.1016/0167-8191(94)90042-6}
  {\path{doi:https://doi.org/10.1016/0167-8191(94)90042-6}}.
\newline\urlprefix\url{https://www.sciencedirect.com/science/article/pii/0167819194900426}

\bibitem{10.1007/978-3-540-85912-3_26}
H.~Haramoto, M.~Matsumoto, P.~L'Ecuyer, {A Fast Jump Ahead Algorithm for Linear
  Recurrences in a Polynomial Space}, in: S.~W. Golomb, M.~G. Parker, A.~Pott,
  A.~Winterhof (Eds.), Sequences and Their Applications - SETA 2008, Springer
  Berlin Heidelberg, Berlin, Heidelberg, 2008, pp. 290--298.

\bibitem{doi:10.1287/ijoc.1070.0251}
H.~Haramoto, M.~Matsumoto, T.~Nishimura, F.~Panneton, P.~L'Ecuyer,
  \href{https://doi.org/10.1287/ijoc.1070.0251}{{Efficient Jump Ahead for
  $\mathbb{F}_2$-Linear Random Number Generators}}, INFORMS Journal on
  Computing 20~(3) (2008) 385--390.
\newblock \href {http://arxiv.org/abs/https://doi.org/10.1287/ijoc.1070.0251}
  {\path{arXiv:https://doi.org/10.1287/ijoc.1070.0251}}, \href
  {https://doi.org/10.1287/ijoc.1070.0251} {\path{doi:10.1287/ijoc.1070.0251}}.
\newline\urlprefix\url{https://doi.org/10.1287/ijoc.1070.0251}

\bibitem{NTL-SHOUP}
{Shoup, Victor}, \href{http://www.shoup.net/ntl}{{NTL}: {A} {L}ibrary for doing
  {N}umber {T}heory, {URL}: {http://www.shoup.net/ntl}} (2021).
\newline\urlprefix\url{http://www.shoup.net/ntl}

\bibitem{Knuth:1997:ACP:270146}
D.~E. Knuth, {The Art of Computer Programming, Volume 2 (3rd ed.):
  Seminumerical Algorithms}, Addison-Wesley, 1997.

\bibitem{lang02}
S.~Lang, Algebra, Springer, New York, NY, 2002.

\bibitem{von_zur_Gathen_Gerhard_2013}
J.~von~zur Gathen, J.~Gerhard, Modern Computer Algebra, 3rd Edition, Cambridge
  University Press, 2013.

\bibitem{10.1145/63039.63042}
S.~K. Park, K.~W. Miller, \href{https://doi.org/10.1145/63039.63042}{Random
  number generators: good ones are hard to find}, Commun. ACM 31~(10) (1988)
  1192–1201.
\newblock \href {https://doi.org/10.1145/63039.63042}
  {\path{doi:10.1145/63039.63042}}.
\newline\urlprefix\url{https://doi.org/10.1145/63039.63042}

\bibitem{MR0044899}
D.~H. Lehmer, Mathematical methods in large-scale computing units, in:
  Proceedings of a {S}econd {S}ymposium on {L}arge-{S}cale {D}igital
  {C}alculating {M}achinery, 1949, Harvard Univ. Press, Cambridge, MA, 1951,
  pp. 141--146.

\end{thebibliography}
\end{document}